\documentclass[12pt,a4paper]{article}

\usepackage[font=scriptsize]{caption}
\usepackage{titlesec}
\usepackage{multicol}
\usepackage{flushend}
\usepackage{amsfonts}
\usepackage{setspace}
\usepackage{balance}
\usepackage{nomencl}
\usepackage{graphicx}
\usepackage[numbers]{natbib}
\usepackage{amsmath}

\usepackage{amsthm}
\usepackage{amsmath}
\usepackage{amssymb}
\usepackage{stmaryrd}
\usepackage{stackrel}
\usepackage{multicol}
\usepackage{url}
\usepackage{color,hyperref}
\usepackage{subfig}
\usepackage{amsmath}
\begin{document}



\title{Thermodynamic Properties of Regular Phantom Black Hole}

\maketitle


\begin{center}
\author{{M. Haditale}, {B. Malekolkalami}}

\thanks{Faculty of Science, University of Kurdistan, Sanandaj, P. O. Box 416, Iran (email:
Maryam.haditale@uok.ac.ir, B.Malakolkalami@uok.ac.ir)}
\end{center}

\begin{abstract}
The  Regular Phantom Black Hole (\textbf{RPBH})s  are of  theoretical and observational importance, some of their properties have been studied.
In this work, we study some of  thermodynamical properties as entropy, temperature, ... in three asymptotically spacetimes, that is, flat, de--Sitter (\textbf{dS}) and Anti--de Sitter (\textbf{AdS}).  Many of the RPBH properties, including horizon radius,  are (directly or indirectly) dependent on a scale parameter $b$. Due to the slightly different structure from Schwarzschild--like metrics, the method to express relations between thermodynamical variables requires a new function of the scale parameter. We  also imply the local and global thermodynamic stability  through the Heat Capacity (\textbf{HC}) and Gibbs Energy (\textbf{GB}), respectively.\\
The calculations and graphs show the results, in the flat background, are very similar to Schwarzschild ones. Also, some of the results show that the asymptotically AdS--RPBH  is more compatible with physical laws than the dS and flat  backgrounds.
\end{abstract}
\section{Introduction}\label{sec-1}
Astronomical observations based on the Type Ia Supernova Project collaboration have revealed the fact that our universe is expanding much faster than in the past \cite{Airy1}. This   has also been confirmed by other projects as Cosmic Microwave Background (\textbf{CMB}) measurements \cite{Airy2, Airy3, Airy4} and  studies of the large scale structure \cite{Airy5}.  Because, the acceleration is slowed by gravity, any proposed candidate to explain the acceleration must have a sufficient negative pressure to counterbalance gravity. It is widely believed that what is causing the expansion is a mysterious entity called Dark Energy (\textbf{DE}) \cite{Airy6}.

One of the most  famous proposed candidate for DE  is the cosmological constant. The successful model in this regard is $\Lambda CDM$, the simplest model that provides a  reasonably good account of the properties and behaviors of the universe. However, despite the great welcome of this theory, it faces two challenges coming from both theoretical and observational sides.  The first extraction of vacuum energy from quantum field theory and second equality of a large amount of  DE  and Dark Matter \cite{Airy7}. Existence of  these challenges and motivation to explain the physical nature of  DE and  its origin has caused   a large number of researches and  works to explore other proposed candidates.

Many astrophysics observations illustrated the pressure to density ratio. For example, a model--free data analysis from 172 type Ia supernovae  resulted in a range of \cite{Airy8}. According to the Plank data during some years,  \cite{Airy9}. Using Chandra Telescope data, hot gas analysis in 26 X-ray luminous dynamically relaxed galaxy clusters gives \cite{Airy10}. The data on SNIa from the SNLS3 sample estimates \cite{Airy11}. In fact, Several DE models with ultra-negative mode equations offer better fit with above data \cite{Airy12, Airy13, Airy14, Airy15}. All of these approaches are in favor of the Phantom DE scenario in which the constant state parameter equation is used \cite{Airy16, Airy17}. The fundamental origin of phantom fields is debatable, but they occur naturally in some models of string theory \cite{Airy18}, supergravity \cite{Airy18}, and theories in more than 11 dimensions, such as F-theory \cite{Airy20}.
Because the phantom field is a candidate for dark energy, the phantom black hole show that singularity in this black hole is destroyed by dark energy \cite{Airy21}. Bronnikov and Fabris studied the regular BHs with self--gravitating, static, spherically symmetric phantom scalar fields with arbitrary potentials in vacuum which are free essential singularity known as RPBH \cite{Airy22}. Regularity is not unique to RPBH, and charged or Gaussian BHs can be mentioned as examples.\\
The thermodynamics of BH \cite{Airy23} is the field that seeks to apply the laws of thermodynamics despite the BH event horizon. Since the study of the statistical mechanics of blackbody radiation led to the development of the theory of quantum mechanics, the attempt to understand the statistical mechanics of BHs had a profound effect on the understanding of quantum gravity, which led to the development of the holographic principle \cite{Airy24}. Over the past 30 years, the research has revealed that there is a very deep and fundamental relationship between gravity, thermodynamics, and quantum theory. The cornerstone of this relationship is the BH thermodynamics, where certain rules of BH mechanics seem to be just ordinary laws of thermodynamics that apply to a system containing BHs. In fact, the BH thermodynamic - mainly obtained by classical and semi-classical analyzes - provides much of our current physical insight into the nature of quantum phenomena occurring in strong gravitational fields \cite{Airy25}.\newline
We study the thermodynamic properties of RPBH. In addition, the considered times for RPBH are the asymptotically flat, dS and AdS and their results are compared with Schwarzschild.\\
The paper is organized as follows. In Sect.\ref{sec-2}, the the Regular Phantom Metric and conditions which are needed that the metric represents a BH is introduced. Sect. \ref{sec-3}, the properties of thermodynamic of RPBH including: Entropy, Temperature, ... is studied. The conclusion are given in Sect.\ref{sec-4}.
\section{The Regular Phantom Metric}\label{sec-2}
A convenient action describing a self--gravitating  scalar field with an arbitrary potential $V(\phi)$, can be written as \cite{Airy7, Airy22}:
\begin{equation}
S=\int{\sqrt{-g} dx^{4}\Big(R+\varepsilon {g^{\mu \nu }}{\partial _{\mu }}\phi {\partial _{\nu }}\phi -2V(\phi )\Big)},
\label{eq1}
\end{equation}
where $R$ is the scalar curvature, $\varepsilon =+1$  describes the usual scalar field with positive kinetic energy and $\varepsilon =-1$ corresponds to the phantom field. Considering static spherically symmetric configuration, a general spacetime metric can be written  as:
\begin{equation}
{ds^{2}}=f(r){dt^{2}}-\frac{{dr}^{2}}{f(r)}-{p^{2}}(r)\Big({d\theta ^{2}}+{\sin \theta ^{2}}{d\varphi ^{2}}\Big),
\label{eq2}
\end{equation}
where  $f(r)$ and $p(r)$ are the  metric functions which are determined by the field equations. By variation of the action (1)  and solving the resulting field equations,  the unknown metric functions  can be obtained as \cite{Airy7, Airy22}:
\begin{equation}
f(r)={p^{2}}(r)\left(\frac{c}{{b}^{2}}+\frac{1}{{p}^{2}(r)}+\frac{3M}{{b}^{3}}\left[\frac{br}{{p}^{2}(r)}+\arctan \left(\frac{r}{b}\right)\right]\right),
\label{eq3}
\end{equation}
\begin{equation}
{p^{2}}(r)={r^{2}}+{b^{2}}.
\label{eq4}
\end{equation}
Also the potential  $V(\phi)$ and scalar field $\phi$  take the following forms:
\begin{align}
V(\phi(r))&=-\frac{c}{{{b}^{2}}}\frac{{{p}^{2}}+2{{r}^{2}}}{{{p}^{2}}}-\frac{3M}{{{b}^{3}}}\Big(\frac{3br}{{{p}^{2}}}+\frac{{{p}^{2}}+2{{r}^{2}}}{{{p}^{2}}}\arctan \left(\frac{r}{b}\right)\Big),\nonumber\\& \phi (r)=\sqrt{2}\in \arctan \Big(\frac{r}{b}\Big)+{{\phi }_{0}}.
\label{eq5}
\end{align}
The metric function(\ref{eq3}) includes three parameters ($M$, $c$, $b$) which the first two are integration constants and the third is a (positive) scale parameter. It determines the connecting strength between phantom scalar field and the gravity\rlap.\footnote{For this reason, it is sometimes called a regular parameter.} Dealing with these parameters requires the following considerations:\\
\begin{figure}[]
\centering
\includegraphics[width=14cm]{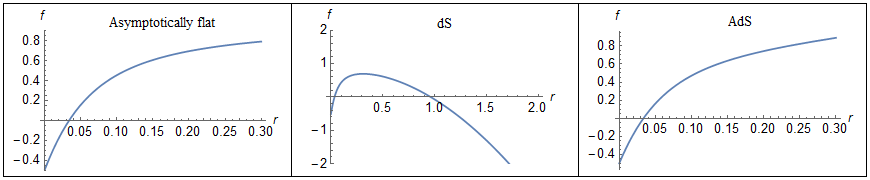}
\caption{The graph of function $f(r)$ (Equation (\ref{eq3})) for $b=0.1$ and $c=-4.5$ for the three asymptotically cases, flat, dS and AdS.}
\label{fig1-1}
\end{figure}
1) A look at the metric function (\ref{eq3}) reveals that a necessary condition to deal with a BH (namely, spacetime including horizon) is
\begin{equation}
c<0.
\label{eq6}
\end{equation}
From here on, we will put $c=-\alpha$ with  $\alpha>0$  and  $\alpha=4.5$ is considered for numerical calculations and graphs.\\
2) The spacetimes described by metric function (\ref{eq3}) includes sixteen classes of possible regular configurations with flat, de Sitter, and anti–de Sitter asymptotics\rlap.\footnote{For a detailed discussion about the bounded values of the current parameters, see \cite{Airy22}.} Corresponding to these three asymptotic cases, there are three bound relations as follows \cite{Airy22}:\\
 for asymptotically flat case:
\begin{equation}
\alpha=\frac{3\pi M}{2b}
\label{eq7}
\end{equation}
and for asymptotically AdS or dS cases:
\begin{equation}
\alpha=-\frac{\pm{2{b}^{3}}-3\pi M}{2b},
\label{eq8}
\end{equation}
which positive (negative) sign corresponds to AdS (dS).\\
3)  In the limit as $b\rightarrow 0$, the metric function (\ref{eq3}) returns to the Schwarzschild metric, therefore,  $M$ is interpreted as the usual mass.\\
When the metric function (3) represents a BH, the horizon radius $r_+$ can be obtained by vanishing the metric function (3), that is
\begin{equation}
f(r_+)=1-\frac{\alpha}{b^2}p^2(r_+)+\frac{3Mr_+}{b^2}+\frac{3M}{b^3} p^2(r_+)\arctan \left(\frac{r_+}{b}\right)=0.
\label{eq9}
\end{equation}
On the other hand, choosing a known value for  $\alpha$, the mass $M$ becomes a function of  $b$ through equation (\ref{eq7}) or (\ref{eq8}).
By taking this in (\ref{eq9}), it turns out that, equation (9)  describes the horizon radius $r_+$ as an implicit function of $b$. The graph of this function is illustrated in Fig.\ref{fig2-2}, for three asymptotically cases as follows:

1) The flat case (left panel): the horizon is linear in $b$. In the next Section, we will see, this linearity  is equal to a scale change in the area of the horizon (or consequently in the entropy) in  compared to the Schwarzschild case.

2) dS case (middle panel): First it is necessary to not that the allowed part of the graph is between two points O and A. This is for the following two reasons:\\
I) The horizon radius is a real  single--valued function.\\
II)  As mentioned, RPBH  $\rightarrow$ Schwarzschild BH as $b\rightarrow 0$  and note that the Schwarzschild BH is asymptotically flat. In other words, the dS  and flat cases are the same as $b\rightarrow 0$.\\
So,  the acceptable part of the graph in Figure 2 indicates that the scale parameter values are limited  to  a domain, that is
\begin{equation}
0<b< b_{max}=b_A\simeq 0.7,
\label{eq10}
\end{equation}
note that the horizon is also restricted between a minimum and a maximum, that is $0<r_+< r_{+max}=r_{+A}$. This means that the dS--RPBH can be formed only for limited values of the scale parameter.\\
3) In AdS case,  the horizon is an increasing function until reaching a maximum (ponit D), then it decreases monotically to an  asymptotic  value (ponit K), that is $\lim_{b\rightarrow \infty}r_+=r_{+K}$. Pay attention here, like in the dS case, there is an upper and lower limit for the horizon radius, with the difference that the scale parameter has no limits.
\begin{figure}
\includegraphics[width=14cm]{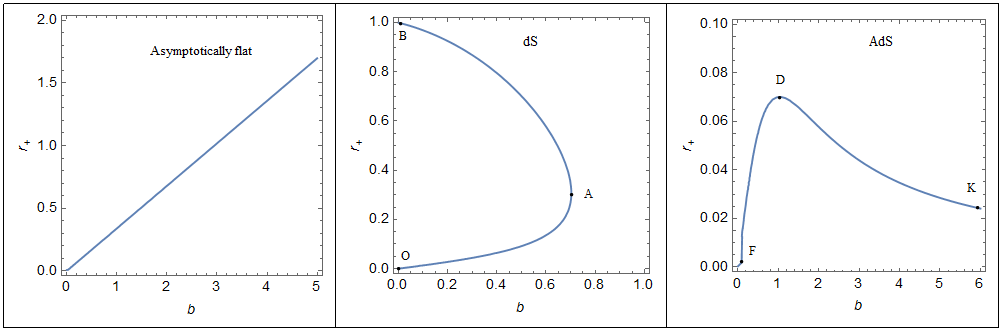}
\caption{The graph of $r_{+}-b$ for three spacetimes.}
\label{fig2-2}
\end{figure}
\subsection{New Parameter}\label{section-2}
In the previous section, the importance of the scale parameter $b$ was discussed to some extent.  As we know, the radius of the horizon is very important in determining the thermodynamical properties of the BH and hence,  it is impossible to  express them  in terms of the scale parameter, because the horizon is an  a implicit function of $b$ (equation (\ref{eq9})).\\
To fix this problem, we introduce  the new parameter defined by:
\begin{center}
$y=\dfrac{r_+}{b},$
\end{center}
which also is  an implicit function of $b$,  however, we will see, it is possible to obtain the inverse (explicit) function, that is $b=b(y)$. This  enable  us  to express the thermodynamical variables  as explicit functions of  $y$ (what is not made possible by $b$). This, in turn, provides the possibility of illustrating thermodynamic diagrams which  play an important role in describing and understanding the thermal properties.\\
To do what was said above, it is necessary to rewrite equation (\ref{eq9}) in the new variable. To do this, we  insert  $M$  from equations (\ref{eq7})  or (\ref{eq8}) into equation (\ref{eq9}), leading to:
\begin{equation}
f(r_+)=g(y)=1-\alpha -\alpha {{y}^{2}}+\frac{2\alpha }{\pi }\lambda \Big(y+(1+{{y}^{2}})\arctan (y)\Big)=0,
\label{eq11}
\end{equation}
where
\begin{equation}
\lambda=\begin{cases}
1, & \text{for  flat case},\\
1\pm\frac{b^2}{\alpha}, & \text{for \textit{AdS}(+) and \textit{dS}(-) cases}.
\end{cases}
\label{eq12}
\end{equation}
For the  flat case, equation (\ref{eq10}) reads
\begin{equation}
g(y)=1-\alpha -\alpha {{y}^{2}}+\frac{2\alpha }{\pi }\left(y+(1+{{y}^{2}})\arctan  (y)\right)=0,
\label{eq13}
\end{equation}
which contains only $y$ and the parameter $\alpha>0$. It is not difficult to verify that to be a real valued $y$, equation (\ref{eq12}) dictates $\alpha>1$. This equivalently means that there isn't any real root  for $0<\alpha<1$\rlap.\footnote{There is only a one root  correspond to  $\alpha=1$, that is $y=0$ which isn't applicable.} Also, there is one to one corresponding between $\alpha$ and $y$ values, for example, for numerical value $\alpha=4.5$, one obtains  $y\approx 1.5$. Thus, an important result   is that, the new parameter has a constant value $y_0$ (depending on $\alpha$), means that  $r_+$ is proportional to $b$, that is
\begin{equation}
r_+=y_0 b,
\label{eq14}
\end{equation}
which confirms Fig.\ref{fig2-2} (left panel).\\
For AdS and dS cases, we   obtain  (from  equation (\ref{eq11}))  $\lambda$  as:
\begin{equation}
\lambda =\frac{\pi }{2}\left(\frac{{{y}^{2}}+1-\frac{1}{\alpha }}{y+(1+{{y}^{2}})\arctan (y)}\right),
\label{eq15}
\end{equation}
now by equating the right hand side (\ref{eq15}) with the right hand side of (\ref{eq12}) (second raw) and arranging the resulting equation in $b^2$, one gets:
\begin{equation}
{{b}^{2}}=\pm \alpha \left(\frac{\pi \Big({{y}^{2}}+1-\frac{1}{\alpha }\Big)}{2\Big(y+(1+{{y}^{2}})\arctan (y)\Big)}-1\right).
\label{eq16}
\end{equation}
The last equation gives the scale parameter as  explicit function of $y$ which  facilitates the expression of thermodynamical  functions in terms of  the new parameter.
\section{Thermal Properties}\label{sec-3}
The goal of this section is the thermodynamical analysis of RPBH. Since, entropy can play a central  role in determining the well--defined thermodynamical quantities, let us examine it first.
\subsection{Entropy}\label{subsec-1}
According to Bekenstein formula, BH entropy is equal to quarter of area bounded by  the event horizon. For  the static and spherically symmetric metric   (\ref{eq2}), the horizon area is $A=4\pi p^2(r_+)$, then   the entropy becomes:
\begin{equation}
S=\frac{A}{4}=\pi \Big({r^{2}+b^{2}}\Big)_{r=r_+},
\label{eq17}
\end{equation}
which in terms of the new parameter  $y=\frac{r_+}{b}$, takes the following form:
\begin{equation}
S=\pi \Big(1+y^2\Big){b}^{2}.
\label{eq18}
\end{equation}
For the next purpose, it is also needful to write the last equation as:
\begin{equation}
S=\pi{r_{+}^{2}}\Big(1+\frac{1}{y^2}\Big).
\label{eq19}
\end{equation}
\begin{center}
\textbf{Flat case}\end{center}
In this case, the new parameter has constant value $y_0$ (equation (\ref{eq14})), thus  equation (\ref{eq18}) reads:
\begin{equation}
S=\Big(1+\frac{1}{y_{0}^{2}}\Big)\pi{r_{+}^{2}}=\Big(1+\frac{1}{y_{0}^{2}}\Big)S_{sch} ,
\label{eq20}
\end{equation}
where $S_{sch}=\pi r_+^2 $ stands for the entropy of the  Schwarzschild BH. Obviously, the entropy of the RPBH is always greater than that of the Schwarzschild BH. As numerical example, for $y_0\approx 1.5$, equation (\ref{eq19}) becomes:
\begin{equation}
S\simeq 1.44S_{sch},
\label{eq21}
\end{equation}
also note that $S\longrightarrow S_{sch}$ as $\alpha\longrightarrow \infty$. Because, from equation (\ref{eq12}) can be easily found that  $y\longrightarrow \infty$ as $\alpha\longrightarrow \infty$.
\begin{center}\textbf{dS(-) and AdS(+) cases}\end{center}
In these cases, by  substituting  $b^2$  from equation  (\ref{eq16})  into (\ref{eq18}), we obtain the entropy as the following explicit  function of $y$:
\begin{equation}
S(y)=\pm \pi\alpha \left( 1+{{y}^{2}} \right) \left(\frac{\pi \Big({{y}^{2}}+1-\frac{1}{\alpha }\Big)}{2\Big(y+(1+{{y}^{2}})\arctan (y)\Big)}-1\right).
\label{eq22}
\end{equation}
As we will see in below, the last equation allows  to plot the graph of the thermodynamic functions versus entropy.
\subsection{Mass}\label{subsec-2}
\begin{figure}[]
\centering
\includegraphics[width=11cm]{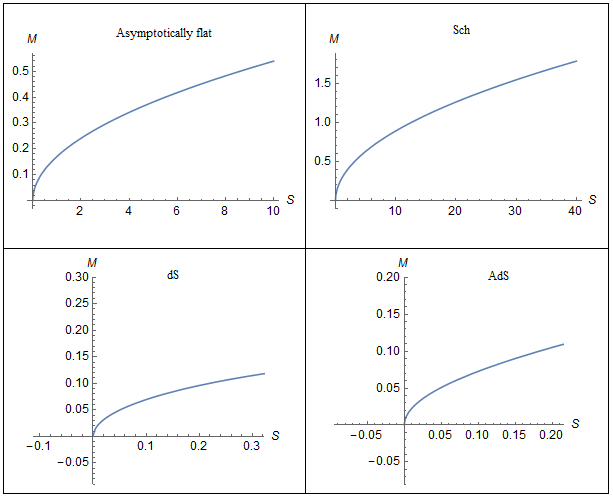}
\caption{The $M-S$ graph, for the flat (\ref{eq24}), dS and AdS (\ref{eq26}) spacetimes in $\alpha=4.5$ and the $M-S$ diagram of Schwarzschild. The results of three spacetimes is similar to Schwarzschild.}
\label{fig3-3}
\end{figure}

In BH thermodynamics, the mass of BH play the role of the internal energy and hence, it is important to consider it from the point of view of system energy changes. To do this, it is common to find the relation between mass and entropy what is presented below.
\begin{center}\textbf{Flat case}\end{center}
In this case, the mass is obtained from (\ref{eq7}), as
\begin{equation}
M=\frac{2\alpha b}{3\pi },
\label{eq23}
\end{equation}
which by substituting $b$ from (\ref{eq18}) reads
\begin{equation}
M=M(S)=\frac{2\alpha}{3 \pi\sqrt{\pi(1+y_0^2)}}  \sqrt{S}\simeq 0.299 \sqrt{S},
\label{eq24}
\end{equation}
where in the last step, we have put $y_0=1.5$.
The variation of mass versus entropy is shown in Fig.\ref{fig3-3} (top left panel). As expected, it is qualitatively similar to the Schwarzschild one (top right panel).\\
\begin{center} \textbf{dS  and AdS cases}\end{center}
In these cases, the mass is obtained from (\ref{eq8}) as:
\begin{equation}
M=\frac{2b}{3\pi }(\pm {{b}^{2}}+\alpha ),
\label{eq25}
\end{equation}
here, contrary to the flat case, it is not possible to obtain the mass as  an explicit function of entropy, however by substituting $b$ from (\ref{eq16}) into equation (\ref{eq25}), it becomes an explicit function of new parameter as:
\begin{equation}
M=M(y)=\frac{2}{3\pi }\sqrt{\pm \left( \frac{\alpha \pi \left( {{y}^{2}}+1-\frac{1}{\alpha } \right)}{2\left( y+\left( 1+{{y}^{2}} \right)\arctan \left[ y \right] \right)}-\alpha  \right)}\left( \frac{\alpha \pi \left( {{y}^{2}}+1-\frac{1}{\alpha } \right)}{2\left( y+\left( 1+{{y}^{2}} \right)\arctan \left[ y \right] \right)} \right).
\label{eq26}
\end{equation}
On other hand, equation (\ref{eq22})  shows the entropy as an explicit function of $y$, that is $S=S(y)$, thus  we can plot the mass variations versus entropy\rlap.\footnote{This can be done, for example by parametricplot code.}  The $M-S$ diagram is illustrated in Fig.\ref{fig3-3} for dS (bottom left panel) and AdS (bottom right panel) cases. They are also Schwarzschild like.
\subsection{Temperature}\label{subsec-3}
\begin{figure}[]
\centering
\includegraphics[width=12cm]{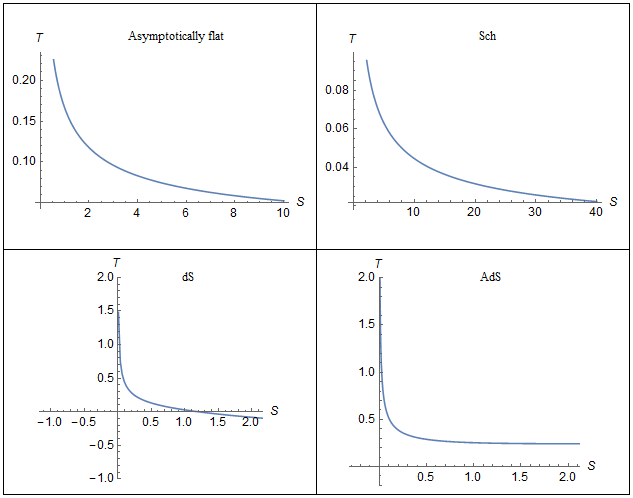}
\caption{The $T-S$ graph, for the flat (\ref{eq31}), dS and AdS (\ref{eq33}) spacetimes in $\alpha=4.5$ and the $T-S$ diagram of Schwarzschild.}
\label{fig4-4}
\end{figure}

For the static, spherically symmetric BH spacetime equipped with metric (\ref{eq2}), the Hawking temperature of the horizon is given by \cite{Airy22}:
\begin{equation}
T=\frac{{f}'(r_+)}{4\pi },
\label{eq27}
\end{equation}
where $f'(r_+)$ denotes the derivative of the metric function  (\ref{eq3}) computed at $r=r_+$. By calculation of derivative of the metric function (\ref{eq3}), the temperature (\ref{eq27}) becomes:
\begin{equation}
T=\dfrac{1}{4 \pi}\left(-\frac{2\alpha}{b} \left(\frac{r_+}{b}\right)+\frac{6M}{{{b}^{2}}}\left(1+\left(\dfrac{r_+}{b}\right)\arctan \left(\frac{r_+}{b}\right)\right)\right),
\label{eq28}
\end{equation}
which  in terms of  $y$, reads
\begin{equation}
T=\left(-\frac{2\alpha}{b}y+\frac{6M}{{{b}^{2}}}\left(1+y\arctan \left(y\right)\right)\right).
\label{eq29}
\end{equation}
In below,  we  plot  the temperature variations versus entropy ($T-S$ diagram), using  (\ref{eq29}).
\begin{center}\textbf{Flat case}\end{center}
In this case, to obtain the temperature, instead of equation (\ref{eq27}), we can use the following simpler formula:\footnote{We note that the  both formulas (27) and (29) lead to the same result.}
\begin{equation}
T=\frac{\partial M}{\partial S},
\label{eq30}
\end{equation}
which by equation (\ref{eq24}) gives:
\begin{equation}
T(S)= \frac{\alpha}{3 \pi\sqrt{\pi(1+y_0^2)S} }  \simeq  \dfrac{0.149}{\sqrt{S}}.
\label{eq31}
\end{equation}
The graph of the function (\ref{eq31}) ($T-S$ diagram)   is shown in Fig.\ref{fig4-4} which is Schwarzschild like, as expected.
\begin{center} \textbf{dS and AdS cases}\end{center}
In these cases, first, by substituting the  mass $M$  from equation  (\ref{eq25}) into equation (\ref{eq29}), we obtain
\begin{equation}
T=\dfrac{1}{4\pi b}\left(-2\alpha y+\dfrac{4}{\pi}(\pm b^2+\alpha)\left(1+y\arctan y\right)\right),
\label{eq32}
\end{equation}
then by inserting (\ref{eq16}) into the last equation, the temperature (\ref{eq29})  becomes a pure function of the new parameter as:
\begin{equation}
T(y)=\frac{-1+\alpha -y\arctan \left( y \right)}{\pi \Big( y+\left( 1+{{y}^{2}} \right)\arctan \left(y \right) \Big)\sqrt{\frac{\mp 4y\alpha \pm 2\pi \left( -1+\alpha +{{y}^{2}}\alpha  \right)\mp 4\left( 1+{{y}^{2}} \right)\alpha \arctan \left(y \right)}{y+\left( 1+{{y}^{2}} \right)\arctan \left(y \right)}}}.
\label{eq33}
\end{equation}
Now, by having (\ref{eq22}), (\ref{eq33}), we can plot $T-S$ diagram   illustrated in Fig.\ref{fig4-4}.
The diagrams   are not Schwarzschild like (contrary to the flat case)\rlap.\footnote{The only similarity between them is that the temperature is  a decreasing function of entropy.} There are two  main differences are:\\
1) In both dS and AdS  cases, the temperature is finite as $S\longrightarrow 0$ while in the flat case, it goes to infinity.\\
2) In dS case,  the temperature  has decreasing nature  and unbounded below, that is it decreases monotonically,  from the high positive values to the high negative values. In AdS case, the temperature is also decreasing, but bounded below, that is  it asymptotes to a certain positive minimum.  In the flat  case,  the temperature asymptotes to zero.

The last difference contains the following physical point: The third law of BH mechanics  states that ''It is not possible to form a BH with vanishing surface gravity (or equivalently vanishing temperature)''. Therefore, the dS--RP BH can violate the third law. The result is that AdS case is more closely related to the physical world.
\subsection{Thermal Stability}\label{section-3}
The HC and GB play an important  role in determining the stability of BHs. They are usually used to analyze the local and global stability of  BHs, respectively. In this subsection, we discuss the stability of the RPBH in the three asymptotically cases.
\subsubsection{Heat Capacity}\label{section-3}
\begin{figure}[]
\centering
\includegraphics[width=14cm]{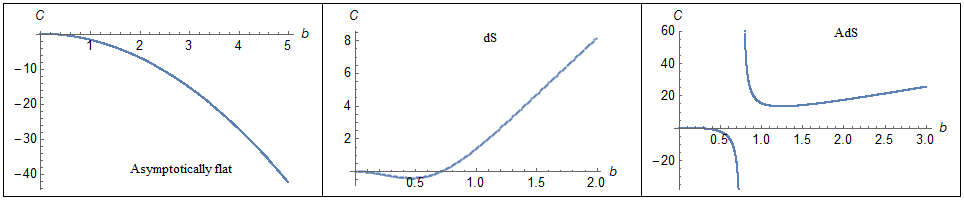}
\caption{The $C-b$ graph, for the flat (\ref{eq38}), dS and AdS (\ref{eq46}) spacetimes in $\alpha=4.5$.}
\label{fig5-5}
\end{figure}

\begin{center}\textbf{Flat case}\end{center}
The HC can be  defined by:
\begin{equation}
C=\frac{dM}{dT},
\label{eq34}
\end{equation}
which  can be written as:
\begin{equation}
C=\frac{dM}{dT}= \frac{\partial M}{\partial S}\frac{dS}{dT}=T{\left(\frac{dT}{dS}\right)^{-1}}.
\label{eq35}
\end{equation}
By equation  (\ref{eq31}), we obtain:
\begin{equation}
\dfrac{dT}{dS}= -\frac{\alpha}{6 \pi\sqrt{\pi(1+y_0^2)}}S^{-3/2},
\label{eq36}
\end{equation}
now by inserting equations (\ref{eq31}) and  (\ref{eq36}) into equation (\ref{eq35}), we get:
\begin{equation}
C(S)=-2S.
\label{eq37}
\end{equation}
This result is the same as HC of the Schwarzschild BH, with the important note, it does not depend on  $\alpha$.  In terms of  scale parameter, HC (\ref{eq37}) becomes:
\begin{equation}
C(b)=-2\pi\Big(1+y_0^2\Big)b^2=-20.41 b^2,
\label{eq38}
\end{equation}
where in the last step we put $y_0=1.5$.\\
The diagram $C-b$ is illustrated  in Fig.\ref{fig5-5} (left panel) which  indicates  the RPBH is locally unstable system for any values of the scale parameter.\\
\begin{center} \textbf{dS and AdS cases}\end{center}
In this case, we rewrite the equation (\ref{eq34}) as:
\begin{equation}
C=\frac{dM}{dT}=\frac{\frac{dM}{db}}{\frac{dT}{db}}
\label{eq39}
\end{equation}
which the numerator can be obtained from equation (\ref{eq25}) as follows:
\begin{equation}
\frac{dM}{db}= \frac{2}{3\pi }(\pm 3{{b}^{2}}+\alpha ).
\label{eq40}
\end{equation}
To calculate the denominator in equation (\ref{eq39}), we can use the following formula:
\begin{equation}
\frac{dT}{db}=\frac{\partial T}{\partial r_+}\frac{\partial r_+}{\partial b}+\frac{\partial T}{\partial b}.
\label{eq41}
\end{equation}
As result by equations (\ref{eq40})(\ref{eq41}), HC (\ref{eq39}) takes the following form:
\begin{equation}
C=\frac{dM}{dT}=\frac{\frac{2}{3\pi }(\pm 3{{b}^{2}}+\alpha )}{\frac{\partial T}{\partial r_+}\frac{\partial r_+}{\partial b}+\frac{\partial T}{\partial b}}.
\label{eq42}
\end{equation}
The three derivatives in the denominator of (\ref{eq41}) can be calculated by equations (\ref{eq28}) and  (\ref{eq9}) as follows:
\begin{equation}
\frac{\partial T}{\partial r_{+}}=-\frac{\alpha -\frac{2\left( \pm {{b}^{2}}+\alpha  \right)\Big( br_{+}+\left( {{b}^{2}}+{{r}^{2}_{+}} \right)\arctan \left[ \frac{r_{+}}{b} \Big] \right)}{\pi \left( {{b}^{2}}+{{r}^{2}_{+}} \right)}}{2\pi{{b}^{2}} },
\label{eq43}
\end{equation}
\begin{equation}
\frac{\partial T}{\partial b}=\frac{\pm 2{{b}^{3}}+2b\alpha -\pi r_{+}\alpha +2r_{+}\left( \pm {{b}^{2}}+\alpha  \right)\arctan \left[ \frac{r_{+}}{b} \right]}{2{{b}^{2}}{{\pi }^{2}}}.
\label{eq44}
\end{equation}
\begin{equation}
\frac{\partial r_{+}}{\partial b}=\frac{r_{+}\Big( \pm {{b}^{3}}\mp 2{{b}^{4}}r+b\alpha -\pi r_{+}\alpha  \Big)+\left( \mp 2{{b}^{4}}+2{{r}^{2}_{+}}\alpha  \right)\arctan \left[ \frac{r_{+}}{b} \right]}{b\Big( \pm {{b}^{3}}\pm 2{{b}^{4}}r_{+}+b\alpha +2{{b}^{2}}r_{+}\alpha -\pi r_{+}\alpha +2r_{+}\left( \pm {{b}^{2}}+\alpha  \right)\arctan \left[ \frac{r_{+}}{b} \right] \Big)}.
\label{eq45}
\end{equation}
Substituting equations (\ref{eq43}) to (\ref{eq45}) into equation  (\ref{eq42}) results:
\begin{align}
C=C(r_+, b)=
(-\frac{\alpha -\frac{2\left( \pm {{b}^{2}}+\alpha  \right)\Big( br_{+}+\left( {{b}^{2}}+{{r}^{2}_{+}} \right)\arctan \left[ \frac{r_{+}}{b} \Big] \right)}{\pi \left( {{b}^{2}}+{{r}^{2}_{+}} \right)}}{2{{b}^{2}}\pi })&\nonumber\\
(\frac{r_{+}\Big( \pm {{b}^{3}}\mp 2{{b}^{4}}r+b\alpha -\pi r_{+}\alpha  \Big)+\left( \mp 2{{b}^{4}}+2{{r}^{2}_{+}}\alpha  \right)\arctan \left[ \frac{r_{+}}{b} \right]}{b\Big( \pm {{b}^{3}}\pm 2{{b}^{4}}r_{+}+b\alpha +2{{b}^{2}}r_{+}\alpha -\pi r_{+}\alpha +2r_{+}\left( \pm {{b}^{2}}+\alpha  \right)\arctan \left[ \frac{r_{+}}{b} \right] \Big)})&\nonumber\\
+(\frac{\pm 2{{b}^{3}}+2b\alpha -\pi r_{+}\alpha +2r_{+}\left( \pm {{b}^{2}}+\alpha  \right)\arctan \left[ \frac{r_{+}}{b} \right]}{2{{b}^{2}}{{\pi }^{2}}})
\label{eq46}
\end{align}
Since $r_+$ is an implicit function of $b$, also  $C$ becomes a function of $b$, that is  $C=C(b)$.  To plot the graph of  this function, we use the relevant  software code and resulting graph  is shown in   Fig.(\ref{fig5-5}) (middle and right panels).\\
The followings can be deduced from the figure:\\
1-- In dS case (middle panel), the RPBH is locally unstable in a certain interval $0<b<b_0$ where $b_0$ is a point in which a  phase transition occurs, that is $C(b_0)=0$. For $b>b_0$,  it seems that the RPBH becomes locally stable, means that,  transition from unstable to stable state  is possible smoothly, but there is a subtle point that should be noted here.  As we saw, in dS case, the scale  parameter values are restricted  to an interval like  (10) with $b_{max}\simeq 0.7$. On the other hand,  as the Fig.(\ref{fig5-5}) (middle panel) shows also $b_0\simeq 0.7$, means that $b_0=b_{max}$. In other words, the allowable part of HC graph is between  $0<b<b_0$ (as horizon graph Fig. 2--middle panel), therefore, the BH  is always unstable.\\
2-- In AdS case (right panel),  the RPBH is locally unstable in a certain interval $0<b<b_0$ where $b_0$ is a point at which a phase transition occurs and for $b>b_0$, the RPBH becomes stable.   But here, the phase transition point is of the infinite discontinuity type, that is $\lim_{b\rightarrow b_0^{\pm}}C(b)=\pm\infty$\rlap.\footnote{The $\pm$  sign on the left hand side stands for left and right limits.} In other words, transition from stable to unstable state (or vice versa) is not possible smoothly and requires an infinite jump. Therefore, depending on the $b$ value, the RPBH is locally always  unstable ($b<b_0$) or always stable ($b>b_0$).
\subsubsection{Gibbs Energy}\label{section-3}
\begin{figure}[!pt]
\centering
\includegraphics[width=14cm]{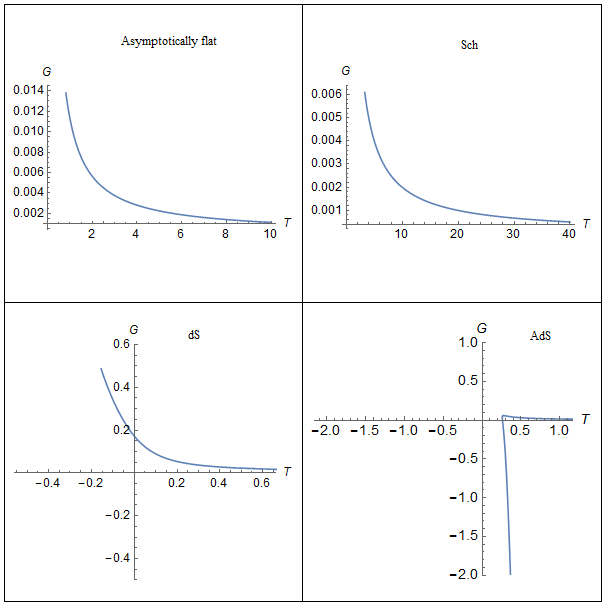}
\caption{The $G-T$ graph, for the flat (\ref{eq48}), dS and AdS (\ref{eq49}) spacetimes in $\alpha=4.5$ and the $G-T$ diagram of Schwarzschild.}
\label{fig6-6}
\end{figure}

To study the global stability of  the BHs, the GE is a useful thermodynamic function \cite{Airy26}.
The BHs are universally stable (unstable) provided their GE is positive (negative)  \cite{Airy27}. Also, in order to investigate and determine the phase transition, it is necessary to calculate GE of the new BHs \cite{Airy28, Airy29, Airy30}. The roots of $G=0$ are the phase transition points \cite{Airy29, Airy31}.\\
The GE formula is defined as:
\begin{equation}
G=M-TS.
\label{eq47}
\end{equation}
To discuss the stability of the BH through  GE, we  study its variations  versus temperature, that is  the $G-T$ diagram.
\begin{center}\textbf{Flat case}\end{center}
By considering the equations (\ref{eq24}), (\ref{eq31}) and (\ref{eq47}), it is not difficult to show that:
\begin{center}
$M=2TS$,
\end{center}
therefore
\begin{equation}
G=TS=\frac{\alpha^2}{9 \pi^3(1+y_0^2) } \dfrac{1}{T},
\label{eq48}
\end{equation}
which behaves  as Homographic function. The graph of this function is  illustrated in Fig.\ref{fig6-6} (top left panel) for $\alpha=4.5$ and it  shows the GE is always positive,  means that  the RPBH is  globally stable. The result is the Schwarzschild like qualitatively (the GE for the Schwarzschild BH is $G_{Sch}= \dfrac{1}{16\pi} \dfrac{1}{T}$). The only difference is the numeric coefficients of $\dfrac{1}{T}$ where in (47), it includes the parameter $\alpha$. This in turn allows to provides adjustment to possible observational data.
\begin{center} \textbf{dS and AdS cases}\end{center}
In these cases, we replace the $S(y)$,  $M(y)$  and $T(y)$  and  in (\ref{eq47}) by (\ref{eq22}), (\ref{eq26}) and (\ref{eq33}) respectively, to obtain  the GE as function of  $y$ as:
\begin{align}
G(y)=\frac{ 1-{{y}^{2}}\left( -3+\alpha  \right)-\alpha +3\left( y+{{y}^{3}} \right)\arctan \left( y \right)}{6\sqrt{2}\Big( y+\left( 1+{{y}^{2}} \right)\arctan \left( y \right) \Big)}\times &\nonumber\\ \sqrt{\frac{\mp 2y\alpha \pm \pi \left( -1+\alpha +{{y}^{2}}\alpha  \right)\mp 2\left( 1+{{y}^{2}} \right)\alpha \arctan \left( y \right)}{y+\left( 1+{{y}^{2}} \right)\arctan \left( y \right)}}.
\label{eq49}
\end{align}
Now by the  parametricplot code,  equations (\ref{eq33}) and (\ref{eq49}) allow to plot  the $G-T$ diagram  illustrated in Fig.\ref{fig6-6} (bottom right and left panels) for $\alpha=4.5$.  The Figure shows in dS case (bottom left panel), the GE is positive and decreasing (asymptotically zero) function of temperature, then the BH is globally stable. In addition, since the first law of  BH thermodynamics prohibits zero temperature, then the maximum stability is at (positive) temperatures close to zero.\\
In AdS case (bottom right panel), in a certain temperature ($T_0 \simeq 0.25$), the GE vanishes and as a result by passing this point undergoes a phase transition ($G=0$). The diagram also shows, the BH temperature can not be smaller than $T_0$. For $T \gtrsim T_0$, GE has positive (small) values and it becomes zero asymptotically with a slowly decreasing rate. Therefore, for this temperature region, the RPBH is globally stable.
\section{conclusion}\label{sec-4}
In this research, the thermodynamics properties of RPBH  are studied and  examined for three asymptotically spacetimes, flat, dS and AdS.
Since many of the characteristics of  RPBH depend on the scale parameter $b>0$ and the Schwarzschild BH is the limiting case of RPBH as $b\rightarrow 0$, the results are compared with the Schwarzschild  BH ones. The results in the asymptotically flat case are   Schwarzschild like, qualitatively.
But in the asymptotically dS and AdS cases, most of the results are somewhat different. It is more significant that the AdS case is more closely related to physical laws. The main conclusions can stated as follows:\\
1) The horizon of the RPBH, in the flat case is a linear increasing function of $b$ . In the dS case, the horizon  is a monastically  increasing function of $b$,  the value of the scale parameter is bounded above $0<b<b_{max}$.  In the AdS case, the scale parameter has no restriction and  the horizon  approaches an asymptotic value (as $b\rightarrow \infty$) which is smaller than the   maximum radius of horizon.\\
2) The entropy of RPBH is always greater than the Schwarzschild one. \\
3) The mass variations versus entropy ($M-S$ diagrams) are Schwarzschild like, in three asymptotically  cases.\\
4) The temperature variations versus entropy ($T-S$ diagram) is  Schwarzschild like in flat case.
The $T-S$ diagram, in dS case displays a decreasing   and unbounded below function which vanishes in a certain entropy violating the third law
of BH thermodynamics. In AdS case, The $T-S$ diagram   displays a decreasing, but bounded
below function which approaches a certain positive minimum asymptotically\rlap.\footnote{Note that in the flat case, this minimum is zero.} Therefore, the AdS--RPBH is more physically acceptable.\\
5)  In the flat case, HC of RPBH is  the same as the Schwarzschild one ($C=-2S$) indicating the locally unstability.\\
In the dS, HC at  allowable interval  ($0<b<b_{max}=b_0$) is negative, hence the RPBH is always unstable.\\
In the AdS case, HC undergoes a phase transition  where the  phase transition point is of an  infinite discontinuity, that is:
\begin{center} $\lim_{b\rightarrow b_0^{\pm}}C(b)=\pm\infty$.\end{center}
Thus, depending on the scale parameter  $b$ is smaller  or greater than the phase transition point $b_0$, the RPBH is always unstable or stable, respectively.\\
6) The graph of the GE  versus temperature $T$ shows in the flat and dS cases, the GE is always positive, indicating the globally stable.
However in the AdS case, the RPBH undergoes a phase transition  at a certain temperature $T_0$, that is $G(T_0)=0$. For  $T \gtrsim T_0$, it becomes globally stable. Also, in this case, the BH temperature    cannot be lower than $T_0$.

\end{document}